# Harnessing cardiac power: heart kinetic motion analysis for energy harvesters

Majid Khazaee[1,*], Milad Hasani[1], Sam Riahi[2,3], Lasse Rosendahl[1], Alireza Rezania[1]

[1] Department of AAU Energy, Aalborg University, Aalborg East, Denmark
[2] Department of Cardiology, Aalborg University Hospital, Aalborg, Denmark
[3] Department of Clinical Medicine, Aalborg University, Aalborg, Denmark

[*] Correspondence author e-mail: mad@energy.aau.dk

**Abstract**

Accurately estimating the complex motion of the heart can unlock enormous potential for kinetic energy harvesting. This paper presents a foundational dataset for heart kinetic motion through in-vivo tests and investigates the most influential factors in heart kinetic motion. In-vivo tests on a living pig's heart, with signal processing, were carried out to study the heart movement by heart beating and respiration motions. A network of nine points on the heart was employed for in vivo measurements. These measurements illustrated the kinetic energy signals in displacement, velocity, and acceleration. The results indicated that the motion level varies in distinct locations over epicardium. The statistical features and autocorrelations were reported for these points, illustrating the highest displacement and acceleration. Each heartbeat generated an energy of 14.35 mJ and a power of 1.03 W. However, this available energy is not uniformly distributed. The results illustrated that not only is cardiac movement location-dependent, but the speed of cardiac displacement cycles is also location-dependent. The right atrium has the highest cardiac kinetic movement with an amplitude of 16.19 mm displacement and 16.3 m/s² acceleration. To evaluate the energy harvesting possibility from the heart's motion, a piezoelectric energy harvester was simulated by the finite element method, implying that the energy harvesting level significantly depends on implant location over epicardium. The results of this study open the potential of designing novel energy harvesters based on accurate heart movements and provide a foundation for future investigations of energy harvesting for leadless pacemaker energy systems.

*Keywords:* Heart Motion, Leadless Pacemaker, Piezoelectric, Energy Harvesting.

## 1. Introduction

Cardiac pacemakers have been implanted for over six decades as a vital intervention for patients with bradycardia (slow heart rate). Over 700,000 patients worldwide receive pacemaker treatment annually [1]. Although the pacemaker device mitigates the dysfunction of the patient's heart, utilizing this device can inversely cause life-threatening complications [2]. The complications associated with pocket and lead of pacemakers could lead to tissue damage and infections [3], [4]. Intra-cardiac leadless Pacemakers (ICLPs) have been developed to tackle some issues of conventional pacemaker devices by eliminating the need for a pocket and leads. Nevertheless, replacing pacemakers due to battery depletion is a critical concern that may expose potential risks to the patient [5]–[8]. In order to reduce the risk of pacemaker lead complications, a new generation of leadless pacemakers has been developed [9]. However, while leadless pacemakers solve numerous complications with leads, they also come with a restricted battery lifespan, necessitating a replacement device every 8-10 years [1], [9], [10]. Prompt replacements are essential to save lives, cut expenses, and alleviate distress. To overcome these limitations, further research is needed to develop novel self-powering ICLPs.

The advent of self-sustaining biomedical implants holds excellent potential for reshaping the field of medicine. In this context, exploring alternative solutions that leverage the kinetic energy the human heart generates is intriguing. Various activities like walking and running can increase the potential kinetic energy harvesting. It is worth noting that the human heart's cardiac cycle consistently produces a mechanical power of 0.93 watts. [11]. Harvesting even a modest fraction of this energy, in the range of 3-5 µW, can potentially trigger a medical revolution. The flexible energy harvesters on the outer wall of the heart, as reported by [12], [13], cannot provide the power for endocardial lead-less pacemakers straightforwardly since the power transfer through heart walls will be challenging and energy-consuming. This paper aims to consider an endocardial energy harvester, which can be integrated with endocardial lead-less pacemaker devices to supply the required electrical energy. In addition to the challenges associated with present energy-scavenging technologies, including inadequate power output [14], weightiness [15], large size [16], and toxicity [17], the precise monitoring of heart motion has been a significant obstacle in the progression of self-sustaining intracardiac implants.

Due to the lack of available heart motion data, accurately estimating the heart-motion energy harvesting in real-time is challenging. Since the heart motion is non-harmonic, complex, and nonuniform [18], researchers often rely on simplified estimates of heart motion. The most common method is harmonic (sinusoidal) vibration simulation [18]–[20]. Another approach is analytical harmonic analysis through Fourier Transform [16]. A reported finite element model of the human heart considers the integrative electromechanical response and includes all four valves and chambers [21]. Researchers have studied heart surface motion in robotic surgery using endoscopic vision systems on porcine 25-kg animals [22]. Direct implantable biocompatible accelerometers have been used to acquire acceleration data from the epicardial wall of a sheep in three axes [23], [24]. Similarly, three-dimensional accelerometers have been used on anaesthetized pigs to measure heart motion near the apex of the heart and coronary artery region, using signal processing to separate respiration and heart movements [25]. Also, in vivo testing of heart motion at two locations around the heart has shown varying peak-to-peak motion [26]. Previous studies on determining heart motion have focused on a single, limited point or reported only acceleration. The assessment has been based on numerical methods when considering the entire heart region.

Integrating energy harvesting technology in leadless pacemakers has significant potential for self-powered cardiac pacemakers. This field has attracted considerable attention [11], [27]–[31]. However, the kinetic analysis of the heart, a critical aspect of energy harvesting from heart motion, is often overlooked. The heart's motion varies across different regions, resulting in different peak-to-peak vibrations across the heart's surface. This variation can affect the output of piezoelectric energy harvesters, which are attached to a single point in the heart that is not a specific point [32]. An urgent need exists for a global map of the heart's motion to accurately predict the potential for energy harvesting across different regions of the heart's surface. Previous in vivo heart motion studies focused on single points or acceleration, and the entire heart region is only assessed by numerical methods [21]. Hence, an unaddressed demand exists for real-time exploration of heart motions at multiple points. This investigation will optimistically benefit the upcoming energy harvesting studies in the heart [11], [13], [30], [31], [33] by providing an accessible database of heart displacement data. It also can be integrated with the studies on novel piezoelectric materials for designing feasibly high-power density energy harvesters [34]–[36]. The present study investigates the heart's physiological kinetic motion induced by breathing and cardiac cycles and assesses the displacement of different points at the pig heart.

This study introduces a heart motion dataset and explores the potential of kinetic energy harvesting across the heart's surface. It covers measurement methods, signal processing techniques for heart displacement signals, and constitutive equations for piezoelectric materials in section 2. Results in section 3 show the nature of heart motion and offer a piezoelectric energy harvester for assessing available kinetic energy. The study sets the foundation for future evaluations of energy harvesting technologies.

## 2. Methods

### 2.1. Heart kinetic motion analysis

Obtaining precise data on the kinetic motion, amplitude, and shape of the human heart is challenging. This study conducts animal surgery tests on living pig hearts to generate a contour map of heart kinetic movement. The pig heart is more similar to human hearts than animal models [37] and can be used to design energy harvesters.

We used measurements obtained through animal surgery to monitor the in vivo kinetic motion of the heart. To achieve this, we used a Doppler laser displacement meter of type ILD1420 from MICRO-EPSION (Ortenburg, Germany) to track the movement of the heart surface. The laser was fixed on a frame facing downwards towards the middle of a beating female pig's heart, and the displacement was measured at a sampling rate of 2000 Hz.

Our study adhered to the requirements of the Danish Animal Experiments Inspectorate for animal treatment, and it was approved by this institution (no. 2021-15-0201-0082). The study animal was a 35 kg female pig that underwent a one-week acclimation period. The Danish Animal Experiments Inspectorate approved the animal facilities and principles of animal care. The pigs were pre-anaesthetized with Zoletil and maintained under anaesthesia with propofol and fentanyl. Bradycardia was treated with Atropin and atrial fibrillation with Amiodaron if needed.

During heart surgery, measurements of the heart's surface were taken in the sagittal plane (frontal plane), as shown in Fig. 1. The measurements were taken at nine different points, covering both the left and right ventricles and right atrium, as shown in Fig. 1 (a). The detailed figure of each measurement point at the animal tests is shown in Supplementary Fig.S1. A laser sensor was placed above the frontal plane and faced downwards, as depicted in Fig. 1 (b). The heart's orientation within the pig's body was at an angle of θ~45°, indicating anterior-posterior motion. The displacement sequence of the heart at point $X$ as $r^X(t_n)$ was due to two primary motions: respiratory and heartbeat motions. The respiratory (breathing) motion has a frequency of less than 0.4 Hz ($f_B$), while the heartbeat motion has a frequency of more than 0.9 Hz ($f_H$) [25]. Frequency analysis is necessary to differentiate between the two motions due to their distinct frequencies. More information can be found in Equation (1.1) and Fig. 1 (c).

$$r^X(t_n) \sim \underbrace{\bar{r}_B^X(t_n)}_{\text{respiratory}} + \underbrace{\bar{r}_H^X(t_n)}_{\text{cardiac cycle}}, n = 1, \dots, N \qquad (1.1)$$

$X$ is the measuring point on the heart surface from $i$ to $ix$.

$$\text{Periodic:} \quad \bar{r}_B^X\left(t_n + \frac{1}{f_B}\right) = \bar{r}_B^X(t_n) \qquad (1.2)$$

$$\text{Periodic:} \quad \bar{r}_H^X\left(t_n + \frac{1}{f_H}\right) = \bar{r}_H^X(t_n) \qquad (1.3)$$

$\bar{r}_H^X$ is the cardiac displacement and $\bar{r}_B^X$ is the respiration displacement function. $f_B$ and $f_H$ are the respiration and cardiac motion frequencies, respectively. The velocity, $v(t)$, and acceleration, $a(t)$, of heart displacement, are obtained by differentiating displacement measurements to time.

The Kurtosis feature measures the sharpness of the cardiac displacement at different points on the heart's surface. A higher Kurtosis value indicates faster cardiac displacement. This feature is calculated in addition to the displacement amplitude.

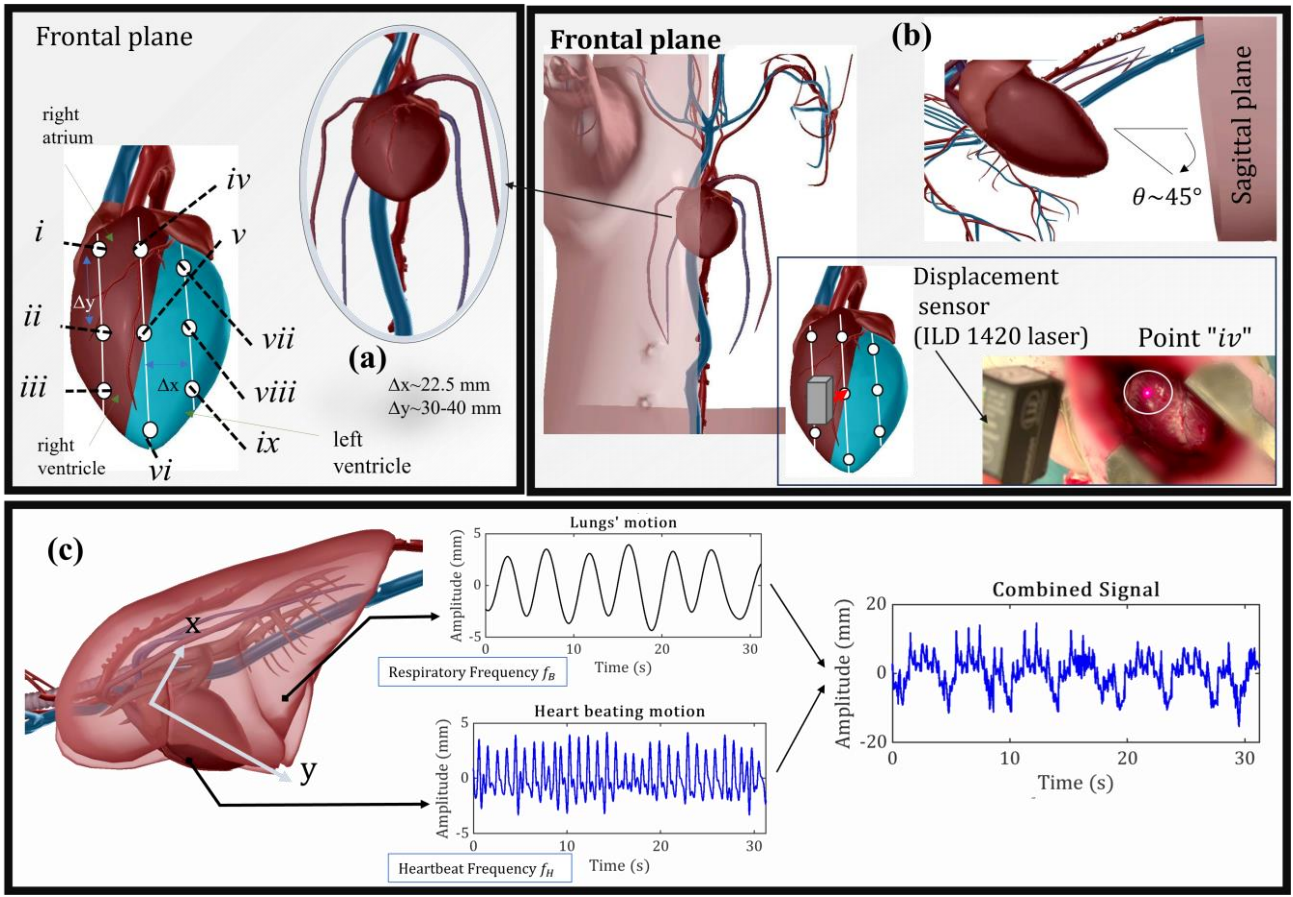

Fig. 1. The measurement of heart kinetic motion in the frontal plane, anterior-posterior motion, (a) the measurement points on the heart surface, (b) the location of the heart and laser sensor, and (c) the two motions of the heart; the lung's motion due to the respiration and the heart beating motion.

We measured the heart's surface displacement using a laser sensor and analyzed the results, as shown in Fig. 2. We removed any bias in the motion data by subtracting the average recorded position. We used the Fast Fourier Transform (FFT) to identify two distinct frequencies related to breathing and heartbeat, as shown in Fig. 2 (b). We separated the motions of breathing and heartbeat by applying bandpass filters with a frequency range of 0.9 to 2.5 Hz for heartbeat bands and 0.1 to 0.4 Hz for breathing bands, as demonstrated in Fig. 2 (c). We further divided the heart motion into individual beats using signal segmentation, as indicated in Fig. 2 (d). Our segmented FFT used a window size of 2000 samples for cardiac cycles and 7500 samples for respiration, with a 75% overlap between segments. Additionally, we calculated the heart rate variation over time using short-term (segmented) FFT, as shown in Fig. 2 (e). This calculation assesses the consistency of heart rate during measurements. We calculated the standard deviation (STD) among the segmented signals.

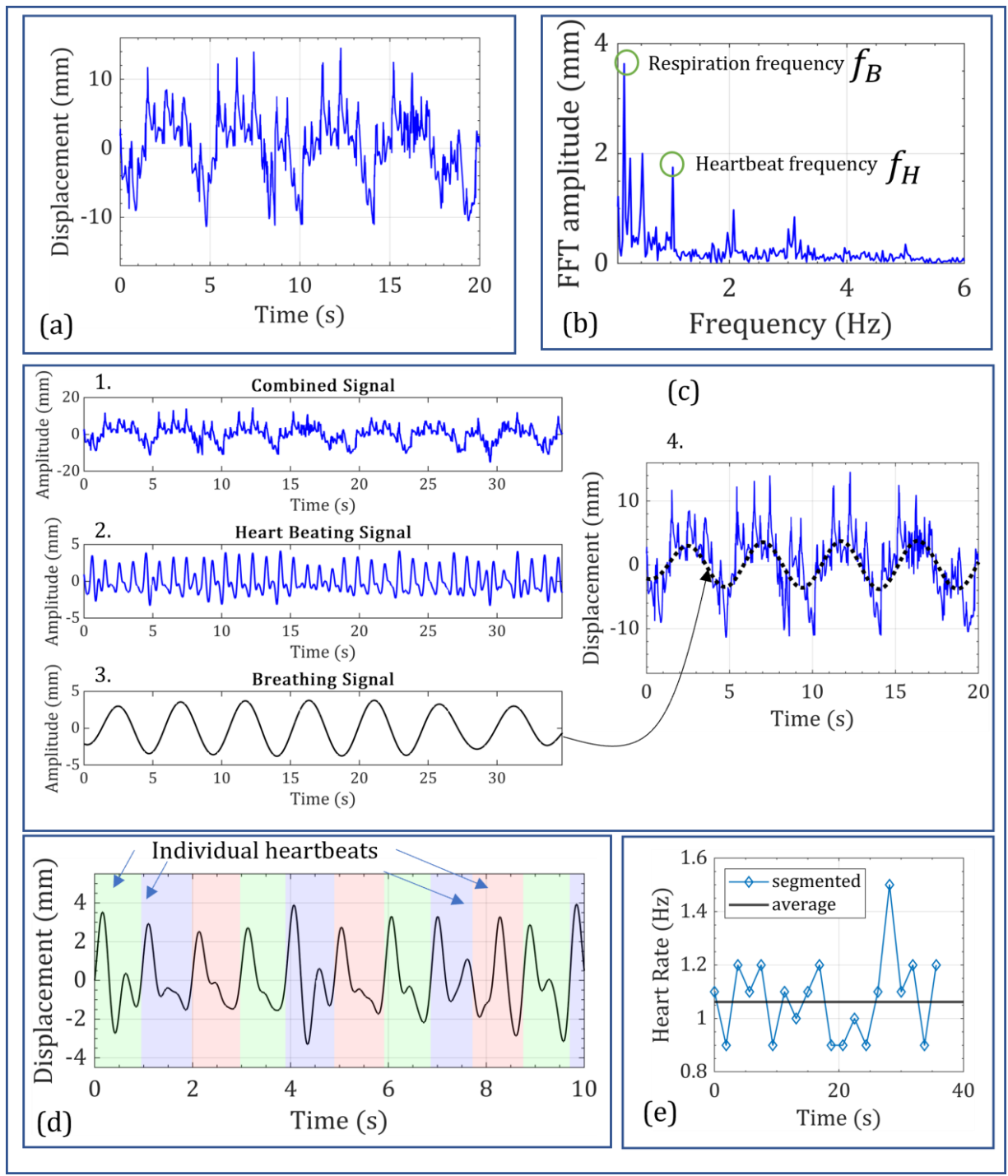

Fig. 2. One sample of the measured heart surface displacement by laser sensor and the postprocessing, (a) unbiased measurement of heart surface displacement, (b) the Fast Fourier Transform (FFT) of the measured displacement, and two distinguished frequencies related to the heartbeat and respiration, (c) the demonstration of separating the heart beating and breathing motions and the combination of these two motions, (d) segmentation of heart beating motion into single individual heartbeats by signal segmentation, and (e) the heart rate variation over the time based on short-term (segmented) FFT.

Autocorrelation and cross-correlation analyses reveal periodic signals in heart displacement measurements obscured by noise from respiration and heartbeat motions. The normalized cross-correlation for the measurement sequences $r^X(t_n) \in \mathbb{R}, n = 1, ..., N$ and $r^Y(t_n) \in \mathbb{R}, n = 1, ..., N$ is $R_{XY}(t_m) \in \mathbb{R}, m = 1, ..., 2N - 1$ and given in Eq. (2).

$$R_{XY}(t_m) = \frac{1}{\sqrt{R_{XX}(0)R_{YY}(0)}} \begin{cases} \sum_{n=0}^{N-m-1} r^X(t_m + t_n)r^Y(t_n), & t_m \geq 0 \\ R_{xx}(-t_m), & t_m < 0 \end{cases} \quad (2)$$

**2.2. Piezoelectricity**

Piezoelectricity refers to the ability of specific materials to produce an electric charge when subjected to mechanical stress or deformation. This property can mainly be found in piezoelectric material of crystalline-based (e.g., lead zirconate titanate) and polymeric-based (e.g., polyvinylidene fluoride). The linear constitutive equations of piezoelectric material can be presented as follows:

$$\begin{aligned} T_{ij} &= c^E_{ijkl}S_{kl} - e_{kij}E_k \\ D_i &= e_{ikl}S_{kl} + \varepsilon^S_{ik}E_k \end{aligned} \quad (3)$$

which $T_{ij}$, $S_{kl}$, $E_k$, and $D_i$ represent the tensors of stress, strain, electric field, and electric displacement, respectively. Moreover, $c^E_{ijkl}$, $e_{kij}$, and $\varepsilon^S_{ik}$ denote the material constants of elastic, piezoelectric, and permittivity, respectively. These equations can analyze piezoelectric materials in different applications, particularly energy harvesting. This study uses the finite element method for analyzing the electromechanical vibration equations of the piezoelectric energy harvesters.

**3. Results and discussions**

This section will present the findings and discussions of the study. Firstly, an analysis of the heart kinetics displacement will be presented, which was observed during in vivo animal testing. Secondly, the role of different kinetic energy points on the heart surface will be demonstrated by evaluating a piezoelectric energy harvester at various locations. This discussion will aid in highlighting the importance of different kinetic energy points on the heart surface.

**3.1. Assessment of heart kinetic motion**

Displayed in Fig. 3 are measurements of heart displacement in the sagittal plane for points "$i$" to "$iv$" on the heart surface, where the heart's kinetic motion is a combination of respiratory and heart-beating motions. Respiratory motion causes the heart to move at a frequency of approximately 0.24 Hz. In contrast, heartbeat motion is a high-amplitude, short-duty cycle half-wave motion and a low-amplitude sinusoidal motion, similar to the description in [22].

Fig. 3 (a) to (d) demonstrate significant variations in heart displacement amplitude at various locations on the heart surface. The total displacement amplitude varies threefold, with some points over 25 mm, while others only ~11 mm. The study's findings are consistent with previous reports of maximum heart motion amplitude of 11 mm [22] and ~10 mm [25], although measured in different regions. This study measured various points of the heart individually to better understand displacement amplitudes at various locations.

The pulse width of the cardiac displacement cycle varies across different points on the heart surface. The pulse width for point "$i$" is half that of point "$ii$". The pulse width is important as it indicates the rate of displacement change, which affects the acceleration of heart regions due to cardiac cycles.

Supplementary Fig.S2 contains heart displacement for other points.

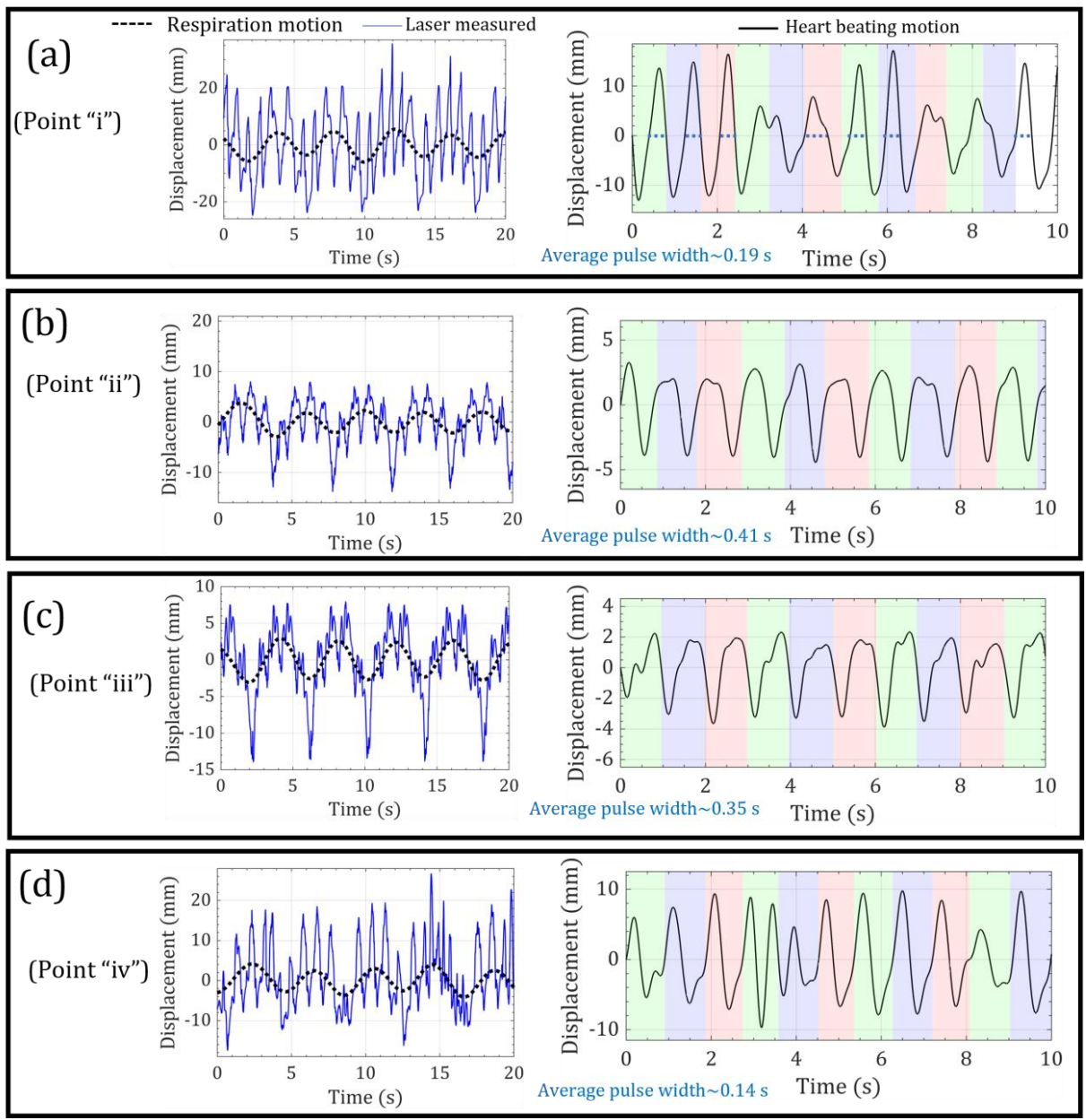

Fig. 3. The laser measurements of heart surface displacements at different points, (a) point "$i$", (b) point "$ii$", (c) point "$iii$", and (d) point "$iv$" with the demonstration of isolated heart beating motion at points "$i$" to "$iv$" on the heart surface.

The short-term FFT of laser displacements can identify the frequencies of respiration and heartbeat motions. Supplementary Fig.S3 shows the FFT signals, which exhibit two peaks for respiration and cardiac-induced motion. Table 1 presents the identified peak frequencies at different points on the heart surface. On average, respiration frequency was 0.25 Hz, and cardiac cycle frequency was 1.14 Hz. However, according to the short-time FFT analysis, the frequency of these motions varied up to 34% for respiration and 19% for the cardiac cycle at one measurement point. This finding is relevant for resonator energy structures sensitive to frequency [16], [38]. Therefore, it is recommended to use broadband piezoelectric harvesters or frequency-tuning designs [39], [40].

Table 1. The frequency of respiration and cardiac motions at different points in the heart is based on short-term FFT.

| The point on the heart | Point description | $f_B$ (Hz) | | $f_H$ (Hz) | |
|---|---|---|---|---|---|
| | | Mean | STD | Mean | STD |
| $i$ | Right atrium | 0.25 | 0.04 | 1.18 | 0.18 |
| $ii$ | Right ventricle | 0.27 | 0.05 | 1.13 | 0.16 |
| $iii$ | Right ventricle | 0.25 | 0.05 | 1.14 | 0.19 |
| $iv$ | Right ventricle outflow tract | 0.25 | 0.07 | 1.19 | 0.22 |
| $v$ | Left ventricle (close to septum between the two ventricles) | 0.24 | 0.04 | 1.15 | 0.22 |
| $vi$ | Left ventricle | 0.25 | 0.06 | 1.10 | 0.18 |
| $vii$ | Left ventricle | 0.25 | 0.09 | 1.15 | 0.19 |
| $viii$ | Left ventricle | 0.24 | 0.06 | 1.14 | 0.21 |
| $ix$ | Left ventricle | 0.26 | 0.06 | 1.12 | 0.18 |
| Average | | 0.25 | 0.01 | 1.14 | 0.03 |

Fig. 4 (a) presents the maximum total heart displacement, velocity, and acceleration at different points in the heart during both the cardiac and respiration cycles. These parameters were measured over multiple heartbeats and kept constant at a heart rate of ~ 60 beats per minute. It is worth noting that there is a significant variation in these kinetic parameters depending on the heart point. The kinetic motion of the heart's surface is not uniformly distributed and differs across various heart regions. Therefore, it is crucial to consider the heart kinetic energy map for energy harvesting assessment. These nonuniform kinetic parameters are closely related to the physiological functioning of the heart. The results indicated that the levels of displacement and acceleration are higher in the right ventricle than in the left ventricle. The right atrium location has the highest acceleration level.

Autocorrelation can be used to study the signals produced by the heart's movement. As shown in Fig. 4 (b), this technique can help identify two periodicities in the signal - one for breathing (respiration) and the other for the heartbeat (cardiac cycle). The influence of these periodicities on the autocorrelation can be observed by looking at the zoomed-in views of the autocorrelation. It is important to note that the effects of respiration and cardiac cycle periodicities vary depending on where the measurement is taken on the heart. For instance, at some points, like "$i$", " the heartbeat periodicity is more noticeable; at other points, like "$iii$," the breathing periodicity has a greater impact than the heartbeat periodicity. Therefore, the overall shape of the autocorrelation is not identical and is affected by cardiac cycle and respiration motions. The autocorrelation shape is directly linked to the acceleration value on the heart point. A more harmonic and uniform autocorrelation corresponds to a lower acceleration level. Moreover, the autocorrelation signals of points "$ii$", "$iii$","$v$", "$vi$", and "$ix$" are similar to the respiration signal, which is more uniform and harmonic (as shown in Fig. 5 (b)). This conclusion indicates that the acceleration level for these points is much lower than the other points.

Walking can add kinetic energy to this acceleration. The acceleration of a man walking at 5 km/h speed was measured at the belly location [41]. The FFT analysis revealed, as shown in Supplementary Figure Fig.S4 that the level of acceleration is below 0.15 m/s$^2$, much lower than the physiological heart motions. Moreover, the peak frequencies are below 7 Hz, which is a low frequency, so the influence on energy harvesting is expected to be low. External activities like walking or running may have additional kinetic effects but affect different heart points uniformly. Considering the uniformity and low level of walking, the energy harvesting performance was solely sought by the physiological heart kinetic motions.

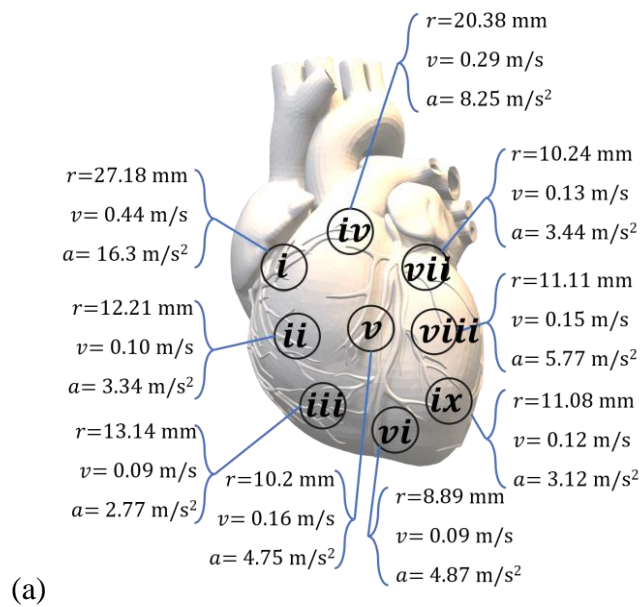

(a)

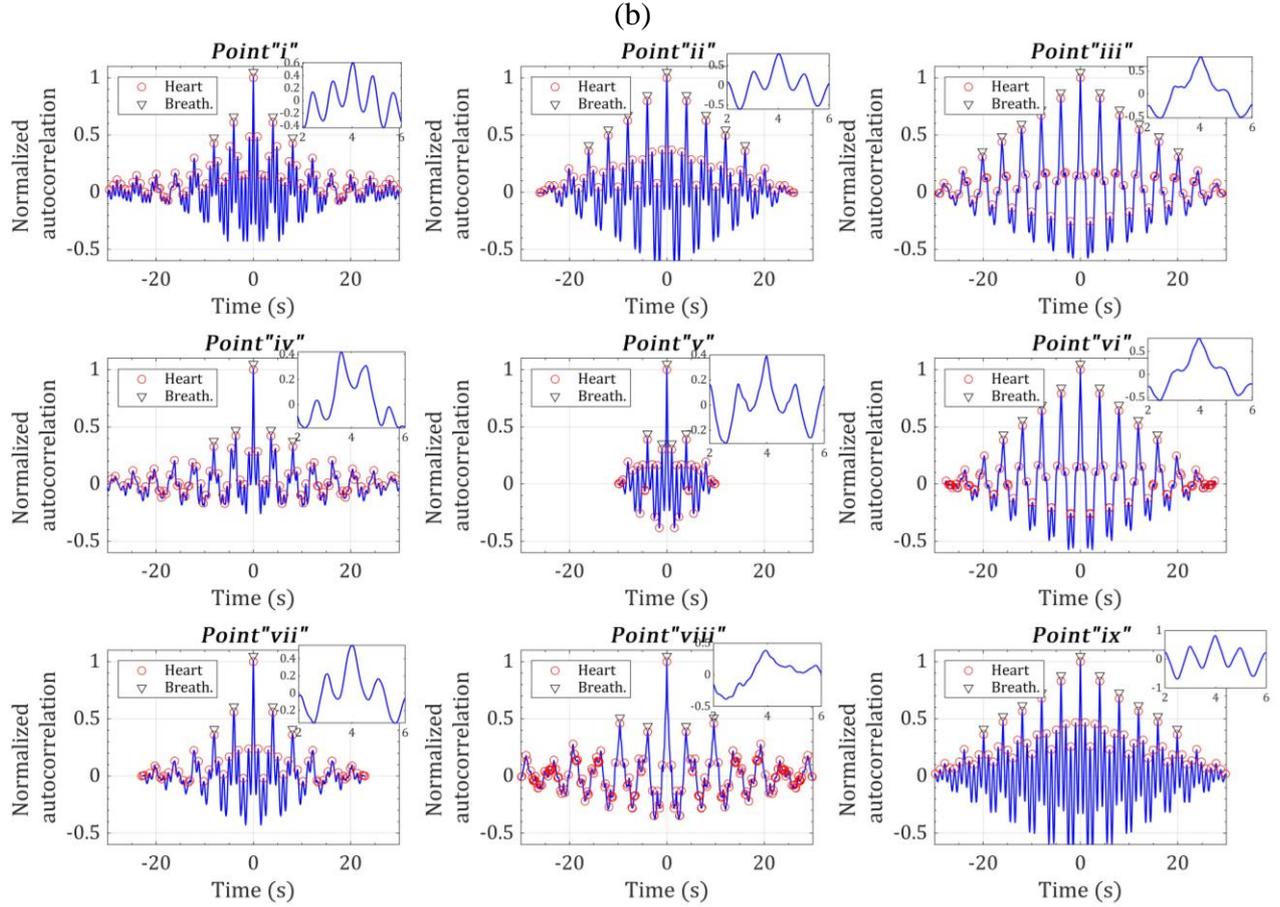

Fig. 4. (a) The maximum displacement, velocity, and acceleration of heart motion include respiration and cardiac cycle, and (b) autocorrelation of cardiac cycle heart displacement at different heart locations.

The heart motion signals due to respiration motion are shown in Fig. 5 (a), while Fig. 5 (b) shows the autocorrelation. The data in Fig. 5 (a) has a consistent pattern across all points with identical autocorrelations in Fig. 5 (b). This similarity is also proven by the high value of cross-correlation factors between the respiration signals, as shown in Table 2.

Table 3 shows that the respiration displacements are harmonics, with an average cross-correlation factor of 0.87 between the respiration signals and a sinusoid standard, indicating a high correlation coefficient between the harmonic and respiration-based heart displacement. The autocorrelation identifies a respiration frequency of 0.24 Hz with STD=0.01 Hz, which agrees well with the short-term FFT output.

In conclusion, the respiration phenomenon creates a harmonic movement at the heart surface, and this harmonic movement is consistent and equally affects the heart surface. Fig. 4 (b) and Fig. 5 (b) show that not all total displacement autocorrelations have a respiration-type pattern. This observation suggests the existence of another signal type linked to cardiac cycle motion, which can be identified through autocorrelation analysis.

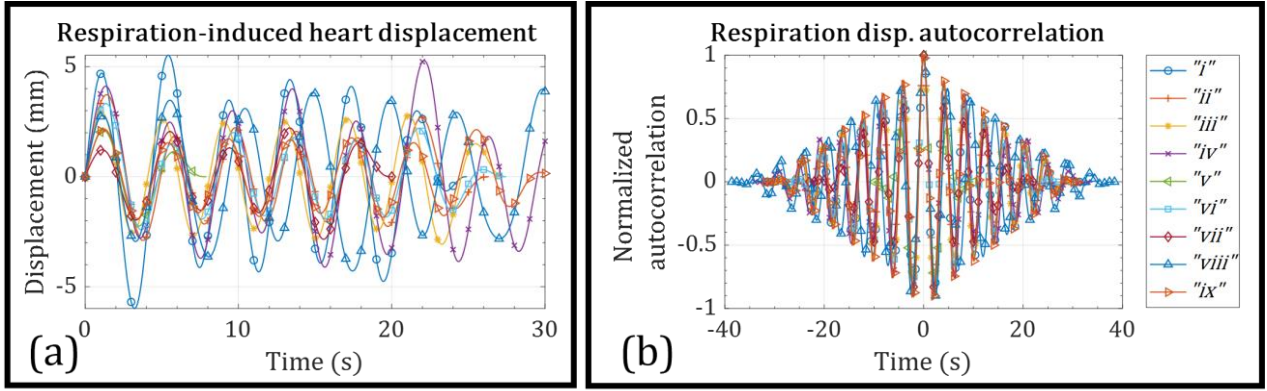

Fig. 5. (a) The respiration-induced heart displacement at different heart surface points and (b) the autocorrelation counterparts.

Table 2. Cross-correlation table for respiration displacement signals

| Point on the heart | i | ii | iii | iv | v | vi | vii | viii | ix |
|---|---|---|---|---|---|---|---|---|---|
| i    | 1.00 | 0.62 | 0.94 | 0.80 | 0.70 | 0.70 | 0.76 | 0.75 | 0.80 |
| ii   | 0.62 | 1.00 | 0.68 | 0.89 | 0.78 | 0.78 | 0.73 | 0.77 | 0.92 |
| iii  | 0.94 | 0.68 | 1.00 | 0.80 | 0.84 | 0.84 | 0.86 | 0.76 | 0.82 |
| iv   | 0.80 | 0.89 | 0.80 | 1.00 | 0.73 | 0.73 | 0.71 | 0.85 | 0.94 |
| v    | 0.70 | 0.78 | 0.84 | 0.73 | 1.00 | 1.00 | 0.85 | 0.80 | 0.78 |
| vi   | 0.70 | 0.78 | 0.84 | 0.73 | 1.00 | 1.00 | 0.85 | 0.80 | 0.78 |
| vii  | 0.76 | 0.73 | 0.86 | 0.71 | 0.85 | 0.85 | 1.00 | 0.80 | 0.80 |
| viii | 0.75 | 0.77 | 0.76 | 0.85 | 0.80 | 0.80 | 0.80 | 1.00 | 0.82 |
| ix   | 0.80 | 0.92 | 0.82 | 0.94 | 0.78 | 0.78 | 0.80 | 0.82 | 1.00 |

Table 3. Cross-correlation factor to a harmonic signal standard

| Point on the heart | Respiration | Cardiac |
|---|---|---|
| i    | 0.92 | 0.66 |
| ii   | 0.75 | 0.81 |
| iii  | 0.93 | 0.80 |
| iv   | 0.87 | 0.17 |
| v    | 0.82 | 0.71 |
| vi   | 0.88 | 0.88 |
| vii  | 0.88 | 0.34 |
| viii | 0.87 | 0.16 |
| ix   | 0.92 | 0.82 |

In the heart's motion of the cardiac cycle, the cross-correlation analysis reveals that there is no similar pattern in cardiac displacement signals, as shown in Table 4. As such, there is a low cross-correlation factor among them. Additionally, the non-normalized autocorrelation of cardiac displacements has been calculated and presented in Supplementary Fig.S5. The autocorrelation of cardiac displacement is not identical to the autocorrelation of a sinusoid. On average, it has a low cross-correlation factor of 0.59 to the sinusoid, as given in Table 2. Some autocorrelation signals in Fig.S5 have an intense high-amplitude peak, like an impulsive signal autocorrelation. This observation proves that the cardiac cycle's motion contains a shock-based motion, which is more prominent at some points than others. This study's results of shock-based heart motion assist the researchers by energy harvesting using shock-based piezoelectric designs [42].

Table 4. Cross-correlation table for cardiac displacement signals

| Point on the heart | $i$ | $ii$ | $iii$ | $iv$ | $v$ | $vi$ | $vii$ | $viii$ | $ix$ |
|---|---|---|---|---|---|---|---|---|---|
| $i$    | 1.00 | 0.53 | 0.52 | 0.46 | 0.57 | 0.57 | 0.52 | 0.53 | 0.59 |
| $ii$   | 0.53 | 1.00 | 0.89 | 0.33 | 0.63 | 0.63 | 0.27 | 0.52 | 0.85 |
| $iii$  | 0.52 | 0.89 | 1.00 | 0.31 | 0.66 | 0.66 | 0.24 | 0.47 | 0.82 |
| $iv$   | 0.46 | 0.33 | 0.31 | 1.00 | 0.41 | 0.41 | 0.37 | 0.43 | 0.35 |
| $v$    | 0.57 | 0.63 | 0.66 | 0.41 | 1.00 | 1.00 | 0.34 | 0.59 | 0.67 |
| $vi$   | 0.57 | 0.63 | 0.66 | 0.41 | 1.00 | 1.00 | 0.34 | 0.59 | 0.67 |
| $vii$  | 0.52 | 0.27 | 0.24 | 0.37 | 0.34 | 0.34 | 1.00 | 0.33 | 0.27 |
| $viii$ | 0.53 | 0.52 | 0.47 | 0.43 | 0.59 | 0.59 | 0.33 | 1.00 | 0.64 |
| $ix$   | 0.59 | 0.85 | 0.82 | 0.35 | 0.67 | 0.67 | 0.27 | 0.64 | 1.00 |

In Fig. 6 (a), the total heart displacement is shown, and the respiratory and cardiac cycle amplitudes are compared. The cardiac cycle amplitude is significantly higher than the amplitude caused by respiration, which averages at 2.87 mm. Furthermore, the cardiac cycle amplitude is nonuniform, and points "$i$" and "$iv$" in the right atrium and right ventricle have much higher displacement than others. The kurtosis feature measures the speed at which displacement cycles occur, which is compared in Fig. 6 (b). The figure shows that respiration has a low kurtosis due to the harmonic type of signal, whereas the cardiac cycle has greater kurtosis than respiration. Fig. 6 (c) displays the cardiac displacement for all nine points on the heart surface. The maximum displacement is at the right atrium, and the minimum is at the middle point "$vi$". Comparing this in vivo cardiac displacement contour with the finite element model of the heart in Ref. [21], it is found that the pattern of having the maximum and minimum amplitudes is identical. However, the maximum cardiac displacement in our in vivo measurement is 16.19 mm, higher than the 10 mm reported in the finite element model [21].

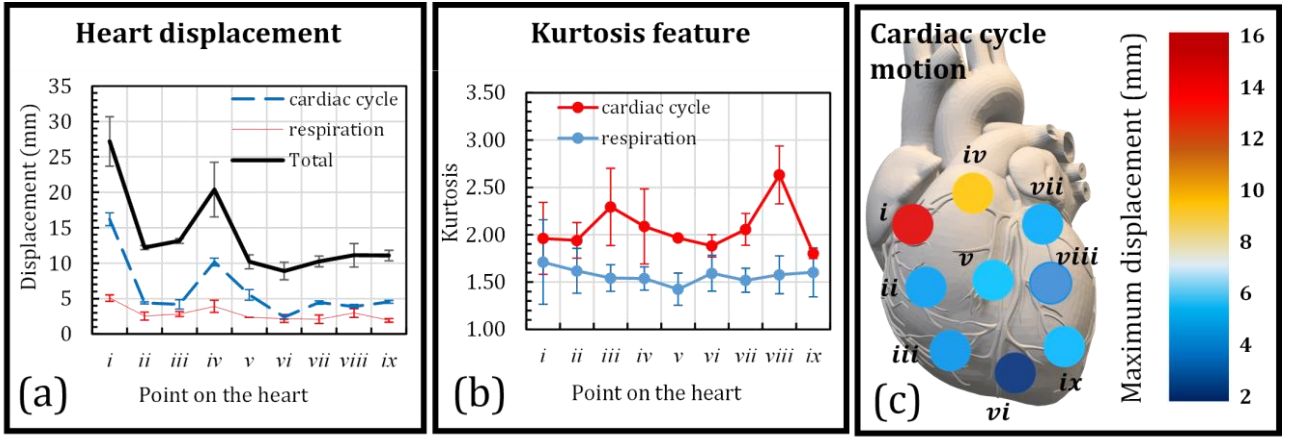

Fig. 6. (a) The maximum displacement, (b) kurtosis feature for the cardiac cycle and respiration motions, and (c) contour of cardiac cycle displacement for the nine points on the heart surface.

Table 5 shows the heart's motion's maximum displacement, velocity, and acceleration. This table considers the movement caused by breathing and the heart's beating. The heart's cycles generate a significant amount of kinetic energy. When evaluating energy harvesting systems, it is crucial to consider the effects of the heart's cycles and other physiological factors on the acceleration of different points on its surface.

Table 5. The maximum displacement, velocity, and acceleration at different points in the heart by considering the heartbeat and breathing motions.

| Point at heart | $r(t_n)$ (mm) | | $v$ (m/s) | | $a$ (m/s$^2$) | | $E_K$ (mJ) | $P_K$ (W) |
|---|---|---|---|---|---|---|---|---|
| | Average | STD | Average | STD | Average | STD | | |
| $i$ | 27.18 | 3.49 | 0.44 | 0.10 | 16.30 | 5.54 | 7.35 | 0.59 |
| $ii$ | 12.21 | 0.30 | 0.10 | 0.02 | 3.34 | 0.59 | 0.36 | 0.02 |
| $iii$ | 13.14 | 0.34 | 0.09 | 0.01 | 2.77 | 0.48 | 0.25 | 0.02 |
| $iv$ | 20.38 | 3.85 | 0.29 | 0.06 | 8.25 | 2.25 | 3.09 | 0.19 |
| $v$ | 10.20 | 0.98 | 0.16 | 0.04 | 4.75 | 0.28 | 1.01 | 0.05 |
| $vi$ | 8.89 | 1.23 | 0.09 | 0.01 | 4.87 | 1.03 | 0.25 | 0.03 |
| $vii$ | 10.24 | 0.77 | 0.13 | 0.02 | 3.44 | 0.60 | 0.57 | 0.03 |
| $viii$ | 11.11 | 1.67 | 0.15 | 0.06 | 5.77 | 2.12 | 1.11 | 0.08 |
| $ix$ | 11.08 | 0.75 | 0.12 | 0.00 | 3.12 | 0.20 | 0.36 | 0.02 |
| Sum | | | | | | | 14.35 mJ | 1.03 W |

An adult pig's heart weighs around 454 grams (397 to 492 grams) [43]. As shown in Table 5, the heart's locations have translational kinetic energy ranges between 0.25 to 7.35 mJ, and the power is between 0.02 to 0.59 W. The summation amount of energy available for translation is 14.35 mJ, and the power generated by the heart is 1.03 W. This calculated power is similar to the previous estimation of 0.93 W [11]. The reported value here is calculated from the in-vivo tests, which are expected to be more practical.

The variation in kinetic energy among different points can significantly impact energy harvesting performance. Therefore, it is crucial to accurately assess the energy harvesting potential at each implant site to determine the optimal location for enhancing energy harvesting performance.

**3.2. Assessment of piezoelectric energy harvesting by heart displacement**

The previous sections assessed the movement over the heart surface to determine the available kinetic energy level on different points. This section attempts to convert the kinetic energy of heart motion to electrical power through piezoelectric energy harvesting. In this regard, a preliminary energy harvester is studied with conviction to highlight the significant impact of the harvester's installation location on its effectiveness. The preliminary endocardial energy harvester with a realistic dimension is considered as Fig. 7 (a), comprising ten spiral piezoelectric beams and tip masses. The dimension of this endocardial energy harvester is identical to the available commercial leadless pacemakers [44]. Each spiral beam comprises a silicon substrate layer sandwiched between two piezoelectric layers made of PZT-5H. These piezoelectric layers are positioned on both sides of the substrate layer with a negative polling direction. More details about the material properties can be found in [45]. The encapsulating PZs with a biocompatible shell (titanium for Micra$^{TM}$ leadless pacemaker, Medtronic, Minneapolis, MN, USA [46]) ensures no harm is inflicted on the surrounding tissue.

A finite element model of the energy harvester is conducted in COMSOL Multiphysics software. A concept of an energy harvester inside a leadless pacemaker is considered, according to Aveir$^{TM}$ leadless pacemaker (Abbott, St. Paul, MN, USA), as shown in Fig. 7 (a). The external shell has a screw-in fixation helix so that the energy harvester is fixed to the heart and vibrates with the heart cardiac cycle. The piezoelectric materials with tip masses have a 0.06 to 0.2 g mass. Therefore, the mass ratio of the leadless pacemaker with the energy harvester is less than 0.9%, and the volumetric ratio is less than 0.4% compared with a female heart [47]. Moreover, the energy harvester is light; the weight is lighter than an identical battery volume, so it does not create adverse effects. In the finite element model, the vibration of each point depicted is individually applied to the bottom side of the proposed energy harvester (Fig. 7 (b)) to derive a steady-state response while the piezoelectric layers are connected in series. For a fair comparison, each point's displacement with an identical period-time of 4 seconds is considered, including four heartbeats and one breathing.

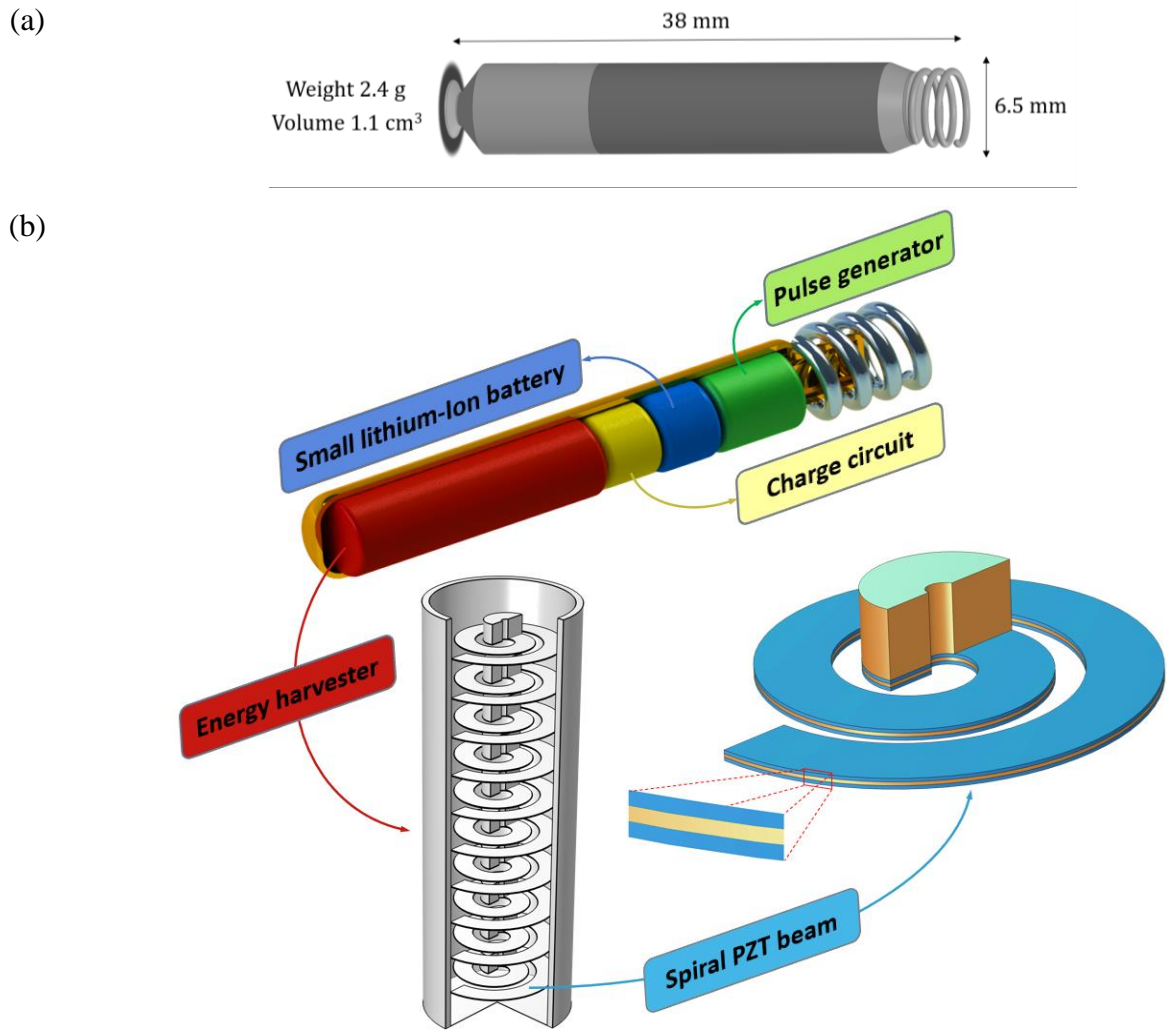

Fig. 7. The schematic of (a) the Aveir[TM] leadless pacemaker (Abbott, St. Paul, MN, USA), (b) the conceptual design of a self-powered pacemaker device with an energy harvester based on ten spiral piezoelectric beams.

The thickness of the substrate ($t_s$) and each piezoelectric ($t_P$) layers can significantly affect the energy harvester output. In this regard, the maximum open-circuit voltage of the energy harvester with different substrate and piezoelectric layers' thickness under motion of point i is shown in Fig. 8 (a). Similarly, the maximum generated power of the energy harvester while connected to resistance 10 kΩ is represented in Fig. 8 (b). This study considers the constraint of a 5 µm minimum threshold for the thickness layer to ensure the durability of the energy harvester against periodic stress during a cardiac cycle. In all the simulations, the maximum stress was below 15 MPa.

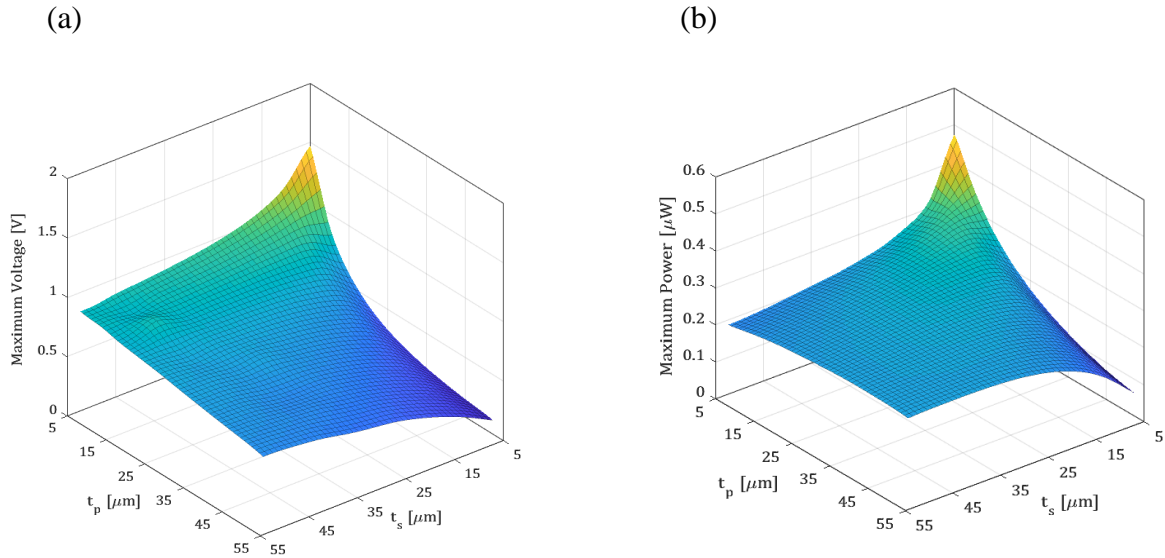

Fig. 8. The outputs of the energy harvester under the motion of point "$i$" during one period: (a) the maximum open-circuit voltage; (b) the maximum generated power while the energy harvester connected to resistance 10 kΩ.

The energy harvester outputs depend on the implant site. Nevertheless, the variation pattern of the maximum open-circuit voltage and power of the energy harvester at other implant points are similar to the results of point "$i$", shown in Fig. 8. For summarization, the normalized generated energy by the energy harvester (with 5 µm substrate and piezoelectric layers thickness) during one cardiac cycle at different implantation points is presented in Fig. 9 (a). The results demonstrate that the harvested energy of point "$i$" dramatically exceeds that of other points, qualifying point "$i$" for endocardial energy harvesting. Moreover, Fig. 9 (b) shows the open-circuit voltage associated with points "$i$" and "$ix$".

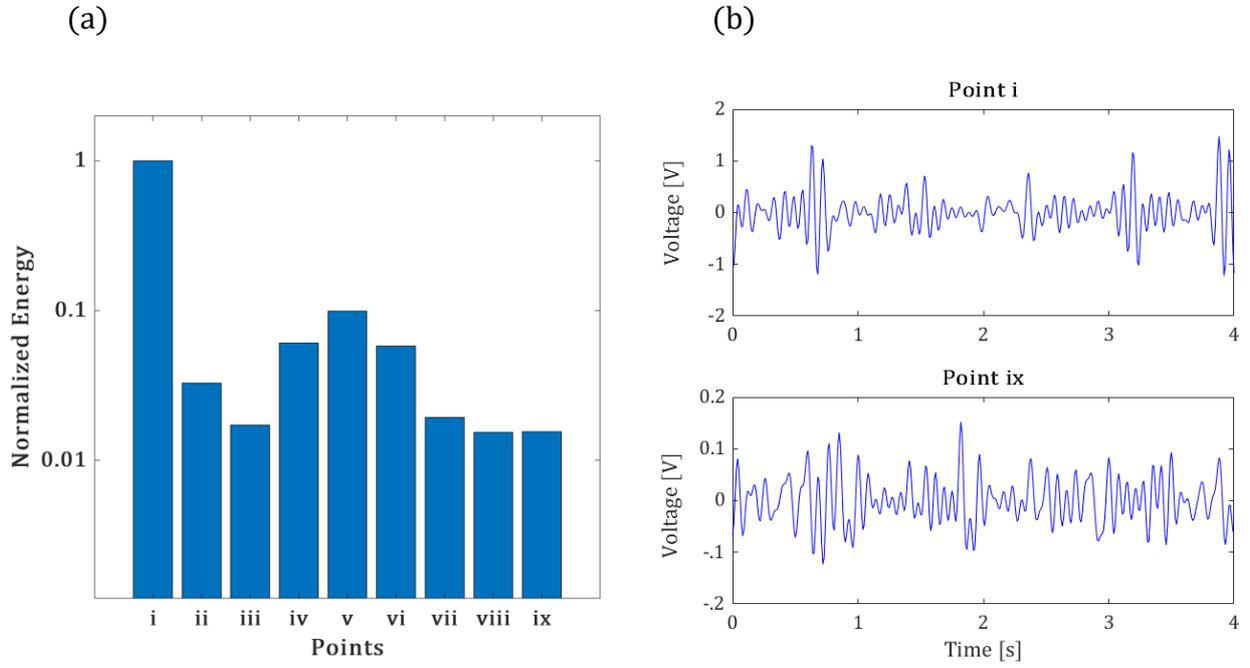

Fig. 9. (a) The normalized periodic harvested energy at each point; (b) one period of generated open-circuit voltage at points "$i$" and "$ix$" for $t_s=t_P=50$ μW.

The results imply that point "$i$" significantly surpasses other points in terms of peak open-circuit voltage, aligning with the remarkable kinetic energy of point "$i$", as detailed in Table 5. Therefore, this location can be one of the nominated sites for implanting the endocardial energy harvester.

**Conclusion**

A foundational dataset for heart kinetic motion through in-vivo tests is presented, investigating the influential factors in heart kinetic motion. In-vivo tests were conducted on a living pig's heart to measure heart and respiration movements. Nine points on the heart surface were used for measurements.

The displacement of the heart can be attributed to two periodic motions: respiration (breathing) and cardiac (heartbeat), which have distinct frequencies. Autocorrelation analysis confirms the periodicity signals from both the cardiac and respiratory motions and indicates their presence at different points on the surface of the heart. The amplitude and frequency of respiration-induced heart motion were recorded to be constant at heart surface points. However, the heart displacement amplitude varies significantly depending on the heart's location due to the cardiac cycle. The right atrium exhibits the highest displacement and acceleration, but the kinetic energy of the left ventricle is low. The right ventricle's top location also has a high cardiac kinetic movement. Our in vivo measurement found that the maximum cardiac displacement was 16.19 mm and acceleration was 16.30 m/s². On average, 14.35 mJ of energy is available for translation, and the power generated by the heart is 1.03 W. The variation in kinetic energy among different points significantly impacts energy harvesting performance. As the piezoelectric model demonstrated, the variation in kinetic energy among different points significantly impacts energy harvesting performance. The autocorrelation proves that the piezoelectric power generation is higher when the periodicity of cardiac-cycle motion is more prominent in the signal.

The current study can initiate several studies. Integrating high-performance designs and novel piezoelectric materials can be optimized for heart displacement measurements, leading to substantial power generation in a feasible dimension and weight. Our team is exploring the in-vivo multi-directional translational and rotational heart measurements and electrophysiological heart models.

**Acknowledgment**

The authors thank Associate Professor Dr. Benedict Kjærgaard from Aalborg University Hospital, Denmark, for his help and support during the animal study. This research is partially financed by the Independent Research Fund Denmark (1031-00001B), the Lundbeck LF-Experiment grant (R324-2019-1747), and the Danish Cardiovascular Academy (funded by the Novo Nordisk Foundation, grant number NNF20SA0067242, and Danish Heart Foundation). This work benefited from the computational resources provided by DeiC National HPC (grant: DeiC-AAU-N2-2023035).

**Competing interests**

The authors declare no competing interests.

**Data Availability**

Real-time in-vivo heart displacement measurements are available.

**References**

[1] S. Sideris *et al.*, "Leadless Cardiac Pacemakers: Current status of a modern approach in pacing," *Hell. J. Cardiol.*, vol. 58, no. 6, pp. 403–410, 2017, doi: 10.1016/j.hjc.2017.05.004.

[2] E. E. Gul and M. Kayrak, "Common Pacemaker Problems: Lead and Pocket Complications," in *Modern Pacemakers*, M. K. Das, Ed., Rijeka: IntechOpen, 2011. doi: 10.5772/12965.

[3] D. Klug *et al.*, "Systemic Infection Related to Endocarditis on Pacemaker Leads," *Circulation*, vol. 95, no. 8, pp. 2098–2107, Apr. 1997, doi: 10.1161/01.CIR.95.8.2098.

[4] K. Shaikhrezai, A. Bartnik, M. Khorsandi, and S. Hunter, "Lead-Sparing Tricuspid Valve Repair Damaged by Pacemaker Lead," *Ann. Thorac. Surg.*, vol. 103, no. 2, pp. e207–e208, Feb. 2017, doi: 10.1016/j.athoracsur.2016.08.029.

[5] L. Sazgary *et al.*, "Incidence of major adverse cardiac events following non-cardiac surgery,"

*Eur. Hear. Journal. Acute Cardiovasc. Care*, vol. 10, no. 5, pp. 550–558, Jun. 2021, doi: 10.1093/EHJACC/ZUAA008.

[6] C. U. Kumari, A. S. D. Murthy, B. L. Prasanna, M. P. P. Reddy, and A. K. Panigrahy, "An automated detection of heart arrhythmias using machine learning technique: SVM," in *Materials Today: Proceedings*, Elsevier Ltd., 2021, pp. 1393–1398. doi: 10.1016/j.matpr.2020.07.088.

[7] World Health Organization (WHO), "Cardiovascular diseases (CVDs)."

[8] M. Kindermann, B. Schwaab, M. Berg, and G. Fröhlig, "Longevity of dual chamber pacemakers: Device and patient related determinants," *PACE - Pacing Clin. Electrophysiol.*, vol. 24, no. 5, pp. 810–815, 2001, doi: 10.1046/j.1460-9592.2001.00810.x.

[9] Gi. Bencardino, R. Scacciavillani, and M. L. Narducci, "Leadless pacemaker technology: Clinical evidence of new paradigm of pacing," *Rev. Cardiovasc. Med.*, vol. 23, no. 2, 2022, doi: 10.31083/J.RCM2302043.

[10] Abbott, "Instructions for Use Aveir." Abbott, 2022.

[11] N. Jackson, O. Olszewski, C. O'Murchu, and A. Mathewson, "Powering a leadless pacemaker using a PiezoMEMS energy harvester," *Smart Sensors, Actuators, MEMS VIII*, vol. 10246, no. June 2017, p. 102460V, 2017, doi: 10.1117/12.2264437.

[12] C. Dagdeviren *et al.*, "Conformal piezoelectric energy harvesting and storage from motions of the heart, lung, and diaphragm," *Proc. Natl. Acad. Sci.*, vol. 111, no. 5, pp. 1927–1932, 2014, doi: 10.1073/pnas.1317233111.

[13] G. T. Hwang *et al.*, "Self-powered cardiac pacemaker enabled by flexible single crystalline PMN-PT piezoelectric energy harvester," *Adv. Mater.*, vol. 26, no. 28, pp. 4880–4887, 2014, doi: 10.1002/adma.201400562.

[14] C. Dagdeviren *et al.*, "Conformal piezoelectric energy harvesting and storage from motions of the heart, lung, and diaphragm," *Proc. Natl. Acad. Sci. U. S. A.*, vol. 111, no. 5, pp. 1927–1932, 2014, doi: 10.1073/pnas.1317233111.

[15] A. Zurbuchen *et al.*, "Energy harvesting from the beating heart by a mass imbalance oscillation generator," *Ann. Biomed. Eng.*, vol. 41, no. 1, pp. 131–141, 2013, doi: 10.1007/s10439-012-0623-3.

[16] M. H. Ansari and M. A. Karami, "Piezoelectric energy harvesting from heartbeat vibrations for leadless pacemakers," *J. Phys. Conf. Ser.*, vol. 660, p. 012121, 2015, doi: 10.1088/1742-6596/660/1/012121.

[17] M. Gholikhani, H. Roshani, S. Dessouky, and A. T. Papagiannakis, "A critical review of roadway energy harvesting technologies," *Appl. Energy*, vol. 261, no. July 2019, p. 114388, 2020, doi: 10.1016/j.apenergy.2019.114388.

[18] N. Jackson, O. Z. Olszewski, C. O'Murchu, and A. Mathewson, "shock-induced aluminum nitride piezoelectric energy harvester leadless pacemaker," *Sensors Actuators, A Phys.*, vol. 264, pp. 212–218, 2017.

[19] A. Kumar, R. Kiran, V. S. Chauhan, R. Kumar, and R. Vaish, "Piezoelectric energy harvester for pacemaker application: A comparative study," *Mater. Res. Express*, vol. 5, no. 7, 2018, doi: 10.1088/2053-1591/aab456.

[20] N. Jackson, O. Olszewski, A. Mathewson, and C. O. Murchu, "Ultra-low frequency piezomems energy harvester for a leadless pacemaker," *Proc. IEEE Int. Conf. Micro Electro Mech. Syst.*, vol. 2018-Janua, no. January, pp. 642–645, 2018, doi: 10.1109/MEMSYS.2018.8346636.

[21] B. Baillargeon, N. Rebelo, D. D. Fox, R. L. Taylor, and E. Kuhl, "The living heart project: A robust and integrative simulator for human heart function," *Eur. J. Mech. A/Solids*, vol. 48, no. 1, pp. 38–47, 2014, doi: 10.1016/j.euromechsol.2014.04.001.

[22] M. Sauvée, P. Poignet, J. Triboulet, E. Dombre, E. Malis, and R. Demaria, "3D Heart Motion Estimation Using Endoscopic Monocular Vision System," *IFAC Proc. Vol.*, vol. 39, no. 18, pp. 141–146, 2006, doi: 10.3182/20060920-3-FR-2912.00029.

[23] L. Brancato, T. Weydts, H. De Clercq, T. Dimiaux, P. Herijgers, and R. Puers, "Biocompatible packaging and testing of an endocardial accelerometer for heart wall motion analysis," *Procedia Eng.*, vol. 120, pp. 840–844, 2015, doi: 10.1016/j.proeng.2015.08.704.

[24] L. Brancato, T. Weydts, W. Oosterlinck, P. Herijgers, and R. Puers, "Biocompatible Packaging of an Epicardial Accelerometer for Real-time Assessment of Cardiac Motion," *Procedia Eng.*, vol. 168, pp. 80–83, 2016, doi: 10.1016/j.proeng.2016.11.152.

[25] L. Hoff, O. J. Elle, M. J. Grimnes, S. Halvorsen, H. J. Alker, and E. Fosse, "Measurements of Heart Motion using Accelerometers," in *The 26th Annual International Conference of the IEEE Engineering in Medicine and Biology Society*, 2004, pp. 2049–2051.

[26] S. Mansouri, F. Farahmand, G. Vossoughi, and A. A. Ghavidel, "A comprehensive multimodality heart motion prediction algorithm for robotic-assisted beating heart surgery," *Int. J. Med. Robot. Comput. Assist. Surg.*, vol. 15, no. 2, Apr. 2019, doi: 10.1002/rcs.1975.

[27] M. Khazaee, S. Riahi, L. Rosendahl, and A. Rezaniakolaei, "Broad-banded frequency-up piezoelectric-based energy harvester from heartbeats' cyclic kinetic motion for leadless pacemakers," in *10th ECCOMAS Thematic Conference on Smart Structures and Materials*, Patras, Greece, 2023.

[28] M. Khazaee, A. A. Enkeshafi, O. Kavehei, S. Riahi, L. Rosendahl, and A. Rezania, "Prospects of self-powering leadless pacemakers using piezoelectric energy harvesting technology by heart kinetic motion," in *45th Annual International Conference of the IEEE Engineering in*

*Medicine and Biology Society*, Sydney: IEEE, 2023.

[29] M. V. Tholl *et al.*, "An intracardiac flow based electromagnetic energy harvesting mechanism for cardiac pacing," *IEEE Trans. Biomed. Eng.*, vol. 66, no. 2, pp. 530–538, 2019, doi: 10.1109/TBME.2018.2849868.

[30] M. Colin, S. Basrour, and L. Rufer, "Design, fabrication and characterization of a very low frequency piezoelectric energy harvester designed for heart beat vibration scavenging," in *Smart Sensors, Actuators, and MEMS VI. Vol. 8763. International Society for Optics and Photonics*, 2013.

[31] M. A. Karami and D. J. Inman, "Powering pacemakers from heartbeat vibrations using linear and nonlinear energy harvesters," *Appl. Phys. Lett.*, vol. 100, no. 4, 2012, doi: 10.1063/1.3679102.

[32] L. V. Boersma *et al.*, "Practical considerations, indications, and future perspectives for leadless and extravascular cardiac implantable electronic devices: a position paper by EHRA/HRS/LAHRS/APHRS," *Europace*, vol. 24, no. 10, pp. 1691–1708, 2022, doi: 10.1093/europace/euac066.

[33] R. Mitra, B. Sheetal Priyadarshini, A. Ramadoss, and U. Manju, "Stretchable polymer-modulated PVDF-HFP/TiO2 nanoparticles-based piezoelectric nanogenerators for energy harvesting and sensing applications," *Mater. Sci. Eng. B Solid-State Mater. Adv. Technol.*, vol. 286, no. September, p. 116029, 2022, doi: 10.1016/j.mseb.2022.116029.

[34] E. J. Curry *et al.*, "Biodegradable nanofiber-based piezoelectric transducer," *Proc. Natl. Acad. Sci. U. S. A.*, vol. 117, no. 1, pp. 214–220, 2020, doi: 10.1073/pnas.1910343117.

[35] E. J. Curry *et al.*, "Biodegradable piezoelectric force sensor," *Proc. Natl. Acad. Sci. U. S. A.*, vol. 115, no. 5, pp. 909–914, 2018, doi: 10.1073/pnas.1710874115.

[36] Y. Jiang, D. Yan, J. Wang, L.-H. Shao, and P. Sharma, "The giant flexoelectric effect in a luffa plant- based sponge for green devices and energy harvesters," *Proc. Natl. Acad. Sci.*, vol. 120, no. 40, p. e2311755120, 2023, doi: 10.1073/pnas.

[37] S. J. Crick, M. N. Sheppard, S. Y. Ho, L. Gebstein, and R. H. Anderson, "Anatomy of the pig heart : comparisons with normal human cardiac structure," *J. Anat.*, vol. 193, no. 1, pp. 105–119, 1998.

[38] S. Shi *et al.*, "Ultra-sensitive flexible piezoelectric energy harvesters inspired by pine branches for detection," *Nano Energy*, vol. 99, no. May, p. 107422, 2022, doi: 10.1016/j.nanoen.2022.107422.

[39] Y. H. Shin *et al.*, "Automatic resonance tuning mechanism for ultra-wide bandwidth mechanical energy harvesting," *Nano Energy*, vol. 77, no. April, p. 104986, 2020, doi: 10.1016/j.nanoen.2020.104986.

[40]  M. A. Halim and J. Y. Park, "Piezoelectric energy harvester using impact-driven flexible sidewalls for human-limb motion," *Microsyst. Technol.*, vol. 24, no. 5, pp. 2099–2107, 2018, doi: 10.1007/s00542-016-3268-6.

[41]  I. Neri, F. Travasso, R. Mincigrucci, H. Vocca, F. Orfei, and L. Gammaitoni, "A real vibration database for kinetic energy harvesting application," *J. Intell. Mater. Syst. Struct.*, vol. 23, no. 18, pp. 2095–2101, 2012, doi: 10.1177/1045389X12444488.

[42]  N. Jackson, O. Z. Olszewski, C. O'Murchu, and A. Mathewson, "Shock-induced aluminum nitride based MEMS energy harvester to power a leadless pacemaker," *Sensors Actuators A Phys.*, vol. 264, pp. 212–218, 2017.

[43]  A. Shah *et al.*, "Anatomical Differences Between Human and Pig Hearts and Their Relevance for Cardiac Xenotransplantation Surgical Technique," *JACC Case Reports*, vol. 4, no. 16, pp. 10–13, 2022, doi: 10.1016/j.jaccas.2022.06.011.

[44]  N. E. Beurskens *et al.*, "Leadless cardiac pacing systems_ current status and future prospects," *Expert Rev. Med. Devices*, vol. 16, no. 11, pp. 923–930, 2019.

[45]  M. Hasani, M. Khazaee, J. E. Huber, L. Rosendahl, and A. Rezania, "Design and analytical evaluation of an impact-based four-point bending configuration for piezoelectric energy harvesting," *Appl. Energy*, vol. 347, no. June, p. 121461, Oct. 2023, doi: 10.1016/j.apenergy.2023.121461.

[46]  "Micra$^{TM}$ MC1VR01 Clinical Manual, Medtronic, Minneapolis, MN, USA." 2021.

[47]  C. E. Friedman and T. Lewis, "The normal size of the heart muscle," *Acta Med. Scand*, vol. 139, pp. 15–19, 1951.